\documentclass[12pt, draftcls, onecolumn]{IEEEtran}
\usepackage{setspace}
\usepackage{multirow}
\usepackage[dvips]{graphicx}
\usepackage[cmex10]{amsmath}
\usepackage{amsmath}
\usepackage{amssymb}
\usepackage{amsfonts}
\DeclareGraphicsExtensions{ps,eps,pst,pdf}
\usepackage{cite}

\newtheorem{theorem}{Theorem}
\newtheorem{lemma}[theorem]{Lemma}
\IEEEoverridecommandlockouts
\begin{document}

\title{Multi-rate Sub-Nyquist Spectrum Sensing in Cognitive Radios}


\author{Hongjian~Sun,~\IEEEmembership{Member,~IEEE,}
 A. Nallanathan$^*$,~\IEEEmembership{Senior Member,~IEEE,}\\
 Jing Jiang,~\IEEEmembership{Member,~IEEE,} and
 Cheng-Xiang Wang,~\IEEEmembership{Senior Member,~IEEE}\\
\thanks{H. Sun and A. Nallanathan$^*$ are with the Department of Electronic Engineering, King's College London, London, WC2R 2LS, UK. (Email: hongjian.sun@kcl.ac.uk; nallanathan@ieee.org)}
\thanks{J. Jiang is with Center for Communication Systems Research, University of Surrey, Guildford, GU2 7XH, UK. (Email: jing.jiang@surrey.ac.uk)}
\thanks{C.-X. Wang is with Joint Research Institute for Signal and Image Processing, School of Engineering~\&~Physical Sciences, Heriot-Watt University, Edinburgh, EH14 4AS, UK. (Email: Cheng-Xiang.Wang@hw.ac.uk) }}

\maketitle

\begin{abstract}
Wideband spectrum sensing is becoming increasingly important to cognitive radio (CR) systems for exploiting spectral opportunities. This paper introduces a novel multi-rate sub-Nyquist spectrum sensing (MS$^3$) system that implements cooperative wideband spectrum sensing in a CR network. MS$^3$ can detect the wideband spectrum using partial measurements without reconstructing the full frequency spectrum. Sub-Nyquist sampling rates are adopted in sampling channels for wrapping the frequency spectrum onto itself. This significantly reduces sensing requirements of CR. The effects of sub-Nyquist sampling are considered, and the performance of multi-channel sub-Nyquist samplings is analyzed. To improve its detection performance, sub-Nyquist sampling rates are chosen to be different such that the numbers of samples are consecutive prime numbers. Furthermore, when the received signals at CRs are faded or shadowed, the performance of MS$^3$ is analytically evaluated. Numerical results show that the proposed system can significantly enhance the wideband spectrum sensing performance while requiring low computational and implementation complexities.

\end{abstract}

\vspace{-0.3cm}


\section{Introduction}
\label{section1}

The radio frequency (RF) spectrum is a scarce natural resource, currently regulated by government agencies. Under the current policy, the primary user (PU) of a particular spectral band has exclusive rights to the licensed spectrum.
With the proliferation of wireless services, the demands for the RF spectrum are continually increasing. On the other hand, it has been reported that localized temporal and geographic spectrum utilization efficiency is extremely low. For example, it has been reported that the maximal occupancy of the spectrum between 30 MHz and 3 GHz is only $13.1\%$ and its average occupancy is $5.2\%$ in New York City~\cite{nsf}.
The spectral under-utilization can be addressed by allowing secondary users to access a licensed band when the PU is absent. Cognitive radio (CR) has become one promising solution for realizing this goal \cite{nallana, chengxiang}.

A crucial requirement of CRs is that they must rapidly fill spectral holes without causing harmful interference to the PUs.
This ability is dependent upon spectrum sensing, which is considered as one of the most critical components in a CR system.
In a multipath or shadow fading environment, the signal-to-noise ratio (SNR) of the primary signal as received at CRs can be severely degraded, which will not only lead to unreliable spectrum sensing results, but will also reduce the capacity of the CR network due to the decreased data transmission time per frame. In such a scenario, cooperative spectrum sensing could increase the reliability of spectrum sensing by exploiting spatial diversity. In our previous work \cite{icc, nallan2}, centralized cooperative spectrum sensing frameworks have been developed for improving the reliability of spectrum sensing. However, these studies only considered narrowband spectrum sensing techniques, the extension to wideband cooperative spectrum sensing requires yet a different approach.

To exploit more spectral opportunities over a large range of frequencies (e.g., 10 kHz $\sim$ 10 GHz), a CR system needs some essential components, i.e., wideband antenna, wideband RF front end, and high speed analog-to-digital converter (ADC). Yoon {\em et al.} \cite{wantenna} have shown that the $-10$~dB bandwidth of the newly designed antenna can be 14.2~GHz. Hao and Hong \cite{wbpf} designed a compact highly selective wideband bandpass filter with a bandwidth of 13.2~GHz. \textbf{In~\cite{LNA}, Bevilacqua and Niknejad designed a wideband CMOS low-noise amplifier with the bandwidth of approximately 10 GHz.} In contrast, the development of ADC technology is relatively behind. To the best of our knowledge, when we require an ADC to have a high resolution and a reasonable power consumption, the achievable sampling rate of the current ADC is 3.6~Gsps~\cite{ADC}. Obviously, ADC becomes a bottleneck in such a wideband system. Even if there exists ADC with more than 20 Gsps sampling rate, the real-time digital signal processing of 20 Gb/s of data could be very expensive.

In previous work, Quan {\em et al.} \cite{quan2, quan} proposed a multiband joint detection (MJD) approach that can sense the primary signal over a wide frequency range. It has been shown that MJD has superior performance for multiband spectrum sensing. In \cite{Tian2006}, Tian and Giannakis studied a wavelet detection approach, which could adapt parameters to a dynamic wideband spectrum. Furthermore, they cleverly introduced compressed sensing (CS) theory to implement wideband spectrum sensing by using sub-Nyquist sampling techniques in the classic paper \cite{scs1}. Later on, the CS-based approach has attracted many talented-researchers' attention in \cite{cite1, cite2, cite3, cite4, GC1, cite5, cite6, ref1, ref2, new1, new2} owning to its advantage of using fewer samples closer to the information rate, rather than the inverse of the bandwidth, to perform wideband spectrum sensing. In \cite{cite1}, Tian {\em et al.} studied cyclic spectrum sensing techniques with high robustness against sampling rate reduction and noise uncertainty. In \cite{cite2}, Zeng {\em et al.} proposed a distributed CS-based spectrum sensing approach for cooperative multihop CR networks.  In our previous work \cite{apwcs, ICC12}, to save system energy, adaptive CS-based spectrum sensing approaches were proposed that could find the best spectral recovery with high confidence.
Unfortunately,  using CS-based approaches, the spectral recovery may cause high computational complexity, leading to a high spectrum sensing overhead due to the restricted computational resources in CRs.

In this paper, we introduce a multi-rate sub-Nyquist spectrum sensing (MS$^3$) approach for cooperative wideband spectrum sensing
in a CR network. Because the spectral occupancy is low, sub-Nyquist sampling is induced in each sampling channel to wrap the
sparse spectrum occupancy map onto itself. The sensing requirements are therefore significantly reduced. We then analyze the effects
caused by sub-Nyquist sampling, and represent the test statistic using a reduced data set obtained from multi-channel sub-Nyquist
sampling. Furthermore, we propose to use different sampling rates in different sampling channels for improving the spectrum sensing performance. Specifically, in the same observation time, the number of samples in multiple sampling channels are chosen as different consecutive prime numbers. In addition, the performance of MS$^3$ for combining faded or shadowed signals is analyzed, and the closed-form bounds for the average probabilities of false alarm and detection are derived. The key advantage of MS$^3$ is that the wideband spectrum can be detected directly from a few sub-Nyquist samples without spectral recovery. Compared to the existing spectrum sensing methods, MS$^3$ can achieve better wideband spectrum sensing performance with a relatively lower implementation complexity.

The rest of the paper is organized as follows. Section \ref{section0} introduces the signal model.
In Section \ref{section2}, we propose the wideband spectrum sensing approach, i.e., MS$^3$.
The performance analysis of MS$^3$ for combining faded signals is given in Section \ref{section3}. Section \ref{section4} presents simulation results, and conclusions are given in Section \ref{section5}.

\section{Preliminary}
\label{section0}

Consider that all CRs keep quiet during the spectrum sensing interval as enforced by protocols, e.g., at the medium access control (MAC) layer \cite{quan2}. Therefore, the observed spectral energy arises only from PUs and background noise.
The bandwidth of the signal as received at CRs is~$W$ (Hertz). Over an observation time $T$, if the sampling rate $f$ ($f \ge 2W$) is adopted to sample the received signal, a sequence of Nyquist samples will be obtained with the length of $JN \stackrel{\triangle}{=} fT$.
This sequence is then divided into $J$ equal-length segments where $N$ denotes the number of Nyquist samples per segment (both $J$ and $N$ are chosen to be natural numbers).
If we use $x_{\textrm{c},i}(t)$ ($t\in [0, T]$) to represent the continuous-time signal received at CR~$i$, after Nyquist sampling, the sampled signal can be denoted by $x_i[n]=x_{\textrm{c},i}(n/f),~ n=0, 1, \cdots, JN-1$. At CR~$i$, the sampled signal of segment $j$ ($j \in [1, J]$) can be written as
\begin{eqnarray}
 x_{i,j}[n]= \left\{ \begin{array}{ll}
x_{\textrm{c},i}(n/f), & n=(j-1)N, (j-1)N+1, \cdots, jN-1\\
0, & \textrm{Otherwise}.
\end{array} \right.
\label{x0}
\end{eqnarray}
The discrete Fourier transform (DFT) spectrum of the sampled signal of segment $j$ is given by
\begin{equation}
X_{i,j}[k]=\sum_{n =0}^{N-1}x_{i,j}[n] e^{-\jmath 2 \pi k n/N},~k=0, 1, \cdots, N-1
\label{spectrum}
\end{equation}
where $\jmath=\sqrt{-1}$. We model spectrum sensing on a frequency bin $k$ as a binary hypothesis test, i.e., $\mathcal{H}_{0,k}$ (absence of PUs) and $\mathcal{H}_{1,k}$ (presence of PUs) \cite{quan}:
\begin{eqnarray}
 X_{i,j}[k]= \left\{ \begin{array}{ll}
Z_{i,j}[k], & \mathcal{H}_{0,k}\\
H_{i,j}[k] S_{i,j}[k]+Z_{i,j}[k], & \mathcal{H}_{1,k}~\textrm{or}~k \in \Omega_i
\end{array} \right.
\label{signal}
\end{eqnarray}
where $Z_{i,j}[k]$ is complex additive white Gaussian noise (AWGN) with zero mean and variance $\delta_{i,k}^2$, i.e., $Z_{i,j}[k]\sim \mathcal{CN}(0, \delta_{i,k}^2)$, $H_{i,j}[k]$ denotes the discrete frequency response between the PU and CR~$i$, $S_{i,j}[k]$ is assumed to be a deterministic signal sent by the PU on the frequency bin $k$, and $\Omega_i$ denotes the spectral support such that $\Omega_i = \{ k | \textrm{PU}~\textrm{presents}~\textrm{at}~X_{i,j}[k] \}$. For simplicity, in the rest of the paper, we assume that the noise variance of the DFT spectrum is normalized to be 1. The observation time $T$ is chosen to be smaller than the channel coherence time so that the magnitude of $H_{i,j}[k]$ remains constant within $T$ for one CR, i.e., constant $|H_{i,j}[k]|$ regarding the segment number $j$.

Because an energy detector does not require any prior information about the transmitted primary signal while having lower complexity than other spectrum sensing approaches \cite{hongjian}, we consider the energy detection approach in this paper. The received signal energy can be calculated as
\begin{equation}
E_i[k]=\sum_{j=1}^{J} \left| X_{i,j}[k] \right|^2, \hspace{1em} k=0, 1, \cdots, N-1.
\label{sum}
\end{equation}
 The decision rule for energy detection approach is then given by \vspace{-1em}
{\setlength{\arraycolsep}{2pt}
\begin{eqnarray}
 & \mathcal{H}_{1, k} & \nonumber \\ \noalign{\vskip -3mm}
 E_i[k] & \gtreqless & \lambda_k, \hspace{1em} k=0,1, \cdots, N-1 \label{decision1} \\ \noalign{\vskip -3mm}
\vspace{-1em} & \mathcal{H}_{0, k} & \nonumber
\end{eqnarray}}
\!\!where $\lambda_k$ is the detection threshold for the frequency bin $k$. Here, it is noteworthy to emphasize that, after Fourier transform, the energy detection is done on each frequency bin in the frequency domain. Thus, the noise in high frequencies should not affect the energy detection in low frequencies, and vice versa. The benefit of frequency-domain energy detection is that the detection performance depends on the SNR \emph{on a single frequency bin, regardless of the noise in the other frequencies} (e.g., high frequency noise due to wideband sensing). To be specific, the signal energy on frequency bin $k$ can be modeled by~\cite{hongjian}
\begin{equation}
 E_i[k]\sim \left\{ \begin{array}{ll}
\chi_{2J}^{2}, \hspace{3em}& \mathcal{H}_{0,k}\\
\chi_{2J}^{2}(2\gamma_i[k]), & \mathcal{H}_{1,k}
\end{array} \right.
\label{ditt}
\end{equation}
where $\gamma_i[k]\stackrel{\triangle}{=} \frac{\mathbb{E}\left( |H_i[k] S_i[k]|^2 \right)}{\delta_{i,k}^2}$ denotes the SNR on the frequency bin $k$ at CR~$i$, $\chi_{2J}^{2}$ denotes central chi-square distribution, and $\chi_{2J}^{2}(2\gamma_i[k])$ denotes non-central chi-square distribution.
 Both of these distributions have $2J$ degrees of freedom and $2\gamma_i[k]$ denotes a non-centrality parameter. Here, the noise has variance $\delta_{i,k}^2$, measured bandwidth $\frac{f}{N}$, and noise temperature $T_n=\frac{\delta_{i,k}^2N}{fK_B}$ where $K_B$ denotes the Boltzmann constant.
The probabilities of false alarm and detection are given by~\cite{hongjian}
\begin{eqnarray}
 P_{\textrm{f},i,k}\!\!&\!\!=\!\!&\!\!\Pr(E_i[k] > \lambda_k |\mathcal{H}_{0,k})= \frac{\Gamma(J,\frac{\lambda_k}{2})}{\Gamma(J)} \\
 P_{\textrm{d},i,k}\!\!&\!\!=\!\!&\!\!\Pr(E_i[k] > \lambda_k |\mathcal{H}_{1,k}) = Q_{J} \left( \sqrt{2\gamma_i[k]}, \sqrt{\lambda_k} \right)
\label{detect2}
\end{eqnarray}
where $\Gamma(a)$ denotes the gamma function, $\Gamma(a, x)$ denotes the upper incomplete gamma function, and $Q_{u}(a,x)$ is the generalized Marcum Q-function defined by $Q_{u}(a,x)=\frac{1}{a^{u-1}}\int_{x}^{\infty}t^{u}e^{-\frac{a^{2}+t^{2}}{2}}I_{u-1}(at)dt$ in which $I_{v}(a)$ is the $v$-th order modified Bessel function of the first kind.


\section{multi-rate sub-Nyquist spectrum sensing}
\label{section2}

It is difficult to realize wideband spectrum sensing, because it requires a high speed ADC for Nyquist rate sampling. We will now present an MS$^3$ system using multiple low-rate samplers to implement wideband spectrum sensing in a CR network.

\subsection{System Description}
\label{section2.1}

\begin{figure}[t]
\centering
\centerline{\includegraphics[width=4.5in]{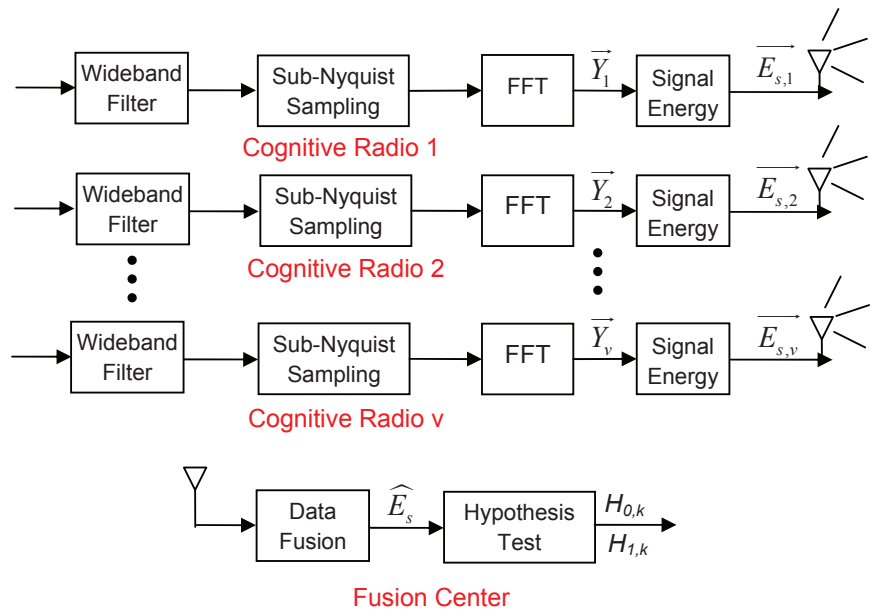}}
\caption{Block diagram of multi-rate sub-Nyquist spectrum sensing (MS$^3$) system.}
\label{fig1}
\end{figure}

Consider that there are $v$ synchronized CRs collaborating for wideband spectrum sensing, and the fusion center (FC) is one of the CRs which has either greater computational resources or longer battery life than other CRs. Due to low spectral occupancy \cite{scs1}, the received signals at CRs are often sparse in the frequency domain.
Here, we assume that the Nyquist DFT spectrum, $\overrightarrow{X_{i,j}} \in \mathbb{C}^N$, is $s$-sparse ($s\ll N$), which means that only the largest $s$ out of $N$ components cannot be ignored. The spectral sparsity level, i.e., $s$, can be obtained from either sparsity estimation~\cite{cite3} or system initialization (e.g., by long term spectral measurements).
As shown in Fig.~\ref{fig1}, MS$^3$ consists of several CRs, each of which has one wideband filter, one low-rate sampler, and a fast Fourier transform (FFT) device. The wideband filters are set to have bandwidth of $W$. MS$^3$ can be described as follows:
\begin{enumerate}
 \item The FC allocates different sub-Nyquist sampling rates to different CRs.
 \item CRs perform sub-Nyquist samplings in the observation time $T$.
 \item The sub-Nyquist DFT spectrum is calculated by using sub-Nyquist samples and FFT device\footnote[1]{Jointly considering wideband spectrum sensing and spectrum reuse in CRs, we use FFT devices for distinguishing different frequencies in order to reuse some un-occupied frequencies. Here, the use of FFT will cause additional complexity of $\mathcal{O}(M\log M)$ and memory storage increment, if $M$ denotes the number of FFT points.}.
 \item The signal energy vectors are formed by using the sub-Nyquist DFT spectrum.
 \item The CRs transmit these signal energy vectors to the FC by using a dedicated common control channel in a band licensed
to the CR network \cite{ccc}.
 \item The received data from all CRs is fused in the FC to form a test statistic.
 \item The FC chooses the detection threshold and performs binary hypothesis tests.
 \item The FC shares the detection results with all CRs.
\end{enumerate}

\subsection{Sub-Nyquist Sampling and Data Combining}
\label{section2.2}

At CR~$i$, we use sub-Nyquist rate $f_i$ ($f_i <2W \le f$) to sample the continuous-time signal $x_{\textrm{c},i}(t)$. The sampled signal can be denoted by
$y_i[n]=x_{\textrm{c},i}(n/f_i),~n=0, 1, \cdots, JM_i-1$ where $JM_i=f_iT$ and $M_i$ is assumed to be a natural number. The sampled signal is then divided into $J$ equal-length segments. The segment $j$ ($j \in [1, J]$) can be written as
\begin{eqnarray}
 y_{i,j}[n]= \left\{ \begin{array}{ll}
x_{\textrm{c},i}(n/f_i), & n=(j-1)M_i, (j-1)M_i+1, \cdots, jM_i-1\\
0, & \textrm{Otherwise}.
\end{array} \right.
\label{x1}
\end{eqnarray}
The DFT spectrum of the sampled signal of segment $j$ ($j \in [1, J]$) can be given by
\begin{equation}
Y_{i,j}[m]=\sum_{n=0}^{M_i-1}y_{i,j}[n] e^{-\jmath 2 \pi m n/M_i},~m=0, 1, \cdots, M_i-1
\label{sspectrum}
\end{equation}
With the aid of Poisson summation formula \cite{poi}, the DFT spectrum of sub-Nyquist samples can be represented by the DFT spectrum of Nyquist samples (as proved in Appendix~\ref{appendix1}):
\begin{equation}
 Y_{i,j}[m]=\frac{M_i}{N} \mathop{\sum}\limits_{l=-\infty}^{\infty} X_{i,j} [m+lM_i], \;\;\; m=0, 1, \cdots, M_i-1.
\label{su3}
\end{equation}
According to (\ref{signal}) and (\ref{su3}), the spectral support of the sub-Nyquist spectrum $\overrightarrow{Y_{i,j}}$ can be given by
\begin{equation}
\Omega_{\textrm{s},i}= \{ m | m=\left|k\right|_{\bmod (M_i)}, \hspace{0.5em} k\in \Omega_i \}.
\label{fold}
\end{equation}

One risk caused by sub-Nyquist sampling is the signal overlap in $Y_{i,j}[m]$. However, when we choose parameters in $JN=fT$ such that $N \gg s$ and let the sub-Nyquist sampling rate satisfy $M_i \sim \mathcal{O}(\sqrt{N})$, the probability of signal overlap is very small (as proved in Appendix~\ref{appendix2}). In such a scenario, we concentrate on considering two cases: no signal on $m$ and one signal on $m$. In the latter case, only a single $l$ is active in (\ref{su3}), and the other terms in the summation of (\ref{su3}) can be modeled as noise by using (\ref{signal}). Thus, the following equation holds from (\ref{signal}) and (\ref{su3}):
\begin{equation}
Y_{i,j}[m]= \frac{M_i}{N} X_{i,j} [ m+lM_i] + \frac{M_i}{N} \sum_{\nu \neq l} Z_{i,j}[m+\nu M_i], \hspace{1em} m+lM_i \in \Omega_i
\label{very}
\end{equation}
where $l$ is an unknown integer within $[0, N/M_i-1]$. Furthermore, using (\ref{signal}) and (\ref{very}), we can model the DFT spectrum of sub-Nyquist samples by
\begin{equation}
 \sqrt{\frac{N}{M_i}}Y_{i,j} \left[\left|k\right|_{\bmod (M_i)} \right]\sim \left\{ \begin{array}{ll}
\mathcal{CN} \bigg(0, \delta_{\textrm{s},i,k}^{2} \bigg), & k \notin \Omega_{i} \\
\mathcal{CN}\left(\sqrt{\frac{M_i}{N}}H_{i,j}[k] S_{i,j}[k], \delta_{\textrm{s},i,k}^{2} \right), & k \in \Omega_{i}
\end{array} \right.
\label{use}
\end{equation}
where $\delta_{\textrm{s},i,k}^{2}$ is the noise variance of sub-Nyquist DFT spectrum, and can be given by using~(\ref{su3})
\begin{equation}
\delta_{\textrm{s},i,k}^{2}= \underbrace{\left\lceil \frac{N}{M_i} \right\rceil }_{No.\;of\; sums} \Bigg( \! \frac{M_i}{N} \!\!\!\! \underbrace{ \sqrt{\frac{N}{M_i}} }_{Scaling\;of\;Y_{i,j}} \!\! \Bigg)^2 \delta_{i,k}^{2} \approx \delta_{i,k}^{2}
\label{noise}
\end{equation}
where $\lceil \frac{N}{M_i} \rceil$ (the smallest integer not less than $\frac{N}{M_i}$) denotes the number of summations in (\ref{su3}).

The signal energy of sub-Nyquist DFT spectrum in each CR node is then calculated by
\begin{equation}
E_{\textrm{s},i}[m]=\sum_{j=1}^{J} \left| Y_{i,j}[m] \right|^2, \hspace{1em} m=0, 1, \cdots, M_i-1.
\label{sum}
\end{equation}
which can be modeled by using (\ref{use}) and (\ref{sum}) as
\begin{equation}
 \frac{N}{M_i}E_{\textrm{s},i}\left[\left|k\right|_{\bmod (M_i)} \right] \sim \left\{ \begin{array}{ll}
 \chi_{2J}^{2}, & k\notin \Omega_{i} \\
 \chi_{2J}^{2} \left( 2\frac{M_i}{N}\gamma_i[k] \right), & k \in \Omega_{i}.
\end{array} \right.
\label{energy3}
\end{equation}
\textbf{We note that, due to the sub-Nyquist sampling, the noise will be folded from the whole bandwidth onto all signals of interest as shown in (\ref{very}). As a result, comparing (\ref{energy3}) with (\ref{ditt}), we find that the received SNR in the sub-Nyquist sampling channel $i$ will degrade from $\gamma_i$ to $\frac{M_i}{N}\gamma_i$. This SNR degradation depends on the ratio between the number of samples at the sub-Nyquist rate and the number of samples at the Nyquist rate (i.e., $\frac{M_i}{N}$).}

\begin{figure}[!ht]
\centerline{\includegraphics[width=4.5in]{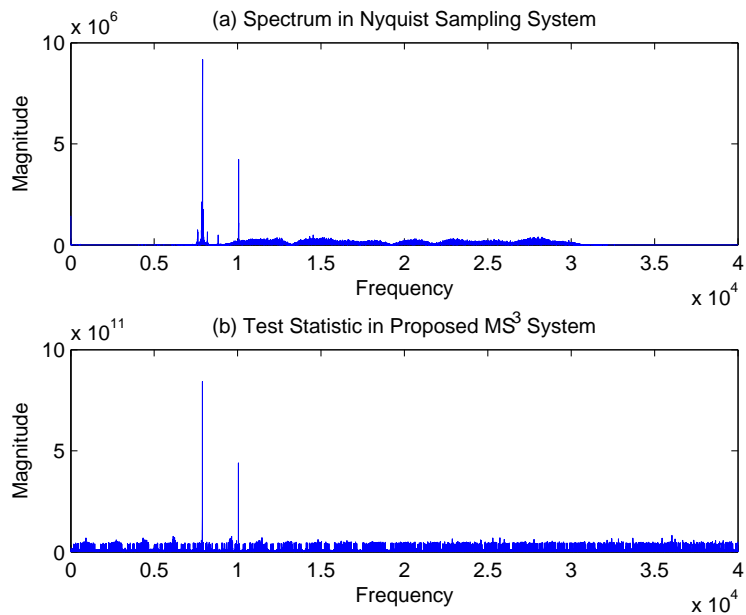}}
\vspace{-1em}
\caption{Illustration of the proposed system for sensing practical ASTC DTV signal: (a) Spectrum of an ASTC DTV signal in Nyquist sampling system, and (b) Test statistic in the proposed system. Here, we use ASTC DTV signal WAS$\_3\_27\_06022000\_$REF and assume the number of CRs $v=22$.  The number of Nyquist samples is $N=80,000$, and the numbers of sub-Nyquist samples in MS$^3$ system are consecutive primes $M_1=1613, M_2=1619, \cdots, M_{22}=1783$. The average compression rate is calculated to $\frac{M}{N}=2.12\%$. }
\label{fig8}
\end{figure}

In MS$^3$, the signal energy vectors at CRs will then be collected at the FC. Finally, we form a test statistic by
\begin{equation}
 \widehat{E_{\textrm{s}}}[k] = \mathop{\sum}\limits_{i=1}^v \frac{N}{M_i}E_{\textrm{s},i}[ |k|_{ \bmod(M_i) }], \hspace{1em} k=0, 1, \cdots, N-1.
\label{energy2}
\end{equation}
In~Fig.~\ref{fig8}, we give an illustration of the above test statistic for a practical ASTC DTV signal.
To test whether the PU is present or not, we adopt the following decision rule: \vspace{-1em}
{\setlength{\arraycolsep}{2pt}
\begin{eqnarray}
& \mathcal{H}_{1, k} & \nonumber \\ \noalign{\vskip -3mm}
\widehat{E_{\textrm{s}}}[k] & \gtreqless & \lambda_k, \hspace{1em} k=0,1, \cdots, N-1. \vspace{-0.6cm} \label{decision} \\ \noalign{\vskip -3mm}
& \mathcal{H}_{0, k} & \nonumber
\end{eqnarray}}\vspace{-2em}

Let $\Omega_{\textrm{A},i}$ denote a set of aliased frequencies (i.e., false frequencies appear as mirror images of the original frequencies around the sub-Nyquist sampling frequency), and $\Omega_{\textrm{U},i}$ represent a set of unaffected/unoccupied frequencies:
\begin{equation}
 \Omega_{\textrm{A},i} \stackrel{\vartriangle}{=}\Big\{ k \Big|m = |k|_{\bmod(M_i)}, \;\; m \in \Omega_{\textrm{s},i}, k \notin \Omega_{i}\Big\}
\label{mirror}
\end{equation}
\begin{equation}
 \Omega_{\textrm{U},i} \stackrel{\vartriangle}{=}\Big\{ k\Big|m=|k|_{ \bmod(M_i) }, \;\; m \notin \Omega_{\textrm{s},i}, k \notin \Omega_{i}\Big\}.
\end{equation}
Using (\ref{energy3}), we can model the test statistic of (\ref{energy2}) as
\begin{equation}
 \widehat{E_{\textrm{s}}}[k] \sim \left\{ \begin{array}{lll}
 \chi_{2Jv}^{2}, & k \in \Omega_{\textrm{U}} \\
 \chi_{2Jv}^{2} \left(\frac{2}{N} \mathop{\sum}\limits_{i\in \Upsilon}^{|\Upsilon|=p} M_i\gamma_i[k] \right), & k \in \Omega_{\textrm{A}}\\
 \chi_{2Jv}^{2}\left(\frac{2}{N}\mathop{\sum}\limits_{i=1}^{v}M_i\gamma_i[k] \right), & k \in \Omega
\label{bound}
\end{array} \right.
\end{equation}
where $\Omega_{\textrm{U}} \stackrel{\vartriangle}{=} \cap_{i=1}^v \Omega_{\textrm{U},i}$, $\Omega_{\textrm{A}} \stackrel{\vartriangle}{=} \cup_{i=1}^v \Omega_{\textrm{A},i}$, $\Omega\stackrel{\vartriangle}{=}\cap_{i=1}^v \Omega_{i}$, $\Upsilon\stackrel{\triangle}{=}\{ i| m = |k|_{\bmod(M_i)}, m \in \Omega_{\textrm{s},i}, k \notin \Omega_{i} \}$ denotes the set of CRs who have aliased frequency on the frequency bin $k$, and $|\Upsilon|=p$ denotes the cardinality of the set $\Upsilon$  (equivalently the number of CRs that have aliased frequencies on the frequency bin $k$).

In (\ref{bound}), $k \in \Omega_{\textrm{U}}$ and $k \in \Omega_{\textrm{A}}$ are two extreme cases under the hypothesis $\mathcal{H}_{0,k}$. The former case denotes there is no aliased frequency on the frequency bin $k$, while the latter case represents there are maximum number of aliased frequencies (i.e., $p$) on the frequency bin $k$. Thus, the former one is the best case while the latter one is the worst case for signal detection under the hypothesis $\mathcal{H}_{0,k}$. The probability of false alarm on the frequency bin $k$ can therefore be bounded by using (\ref{bound})
\begin{eqnarray}
 { \frac{\Gamma(Jv, \frac{\lambda_k}{2})}{\Gamma(Jv)} \le P_{\textrm{f},k} \le Q_{Jv} \left( \sqrt{ \frac{2}{N} \mathop{\sum}\limits_{i\in \Upsilon}^{|\Upsilon|=p} M_i \gamma_i[k] },\sqrt{\lambda_k}\right) }.
\label{ff1}
\end{eqnarray}
We note that the problem of minimizing the probability of false alarm can be transformed to minimize the parameter $p$, which depends on several factors, e.g., the sampling rates of CRs. Using the same sub-Nyquist sampling rates in MS$^3$ is not recommended as it could lead to $p=v$, resulting in the maximum of the probability of false alarm.  As we will see in the following subsection, the parameter $p$ can be minimized by using different sampling rates at CRs.

\subsection{Multi-rate Sub-Nyquist Spectrum Sensing}
\label{section2.3}

To improve the detection performance of sub-Nyquist sampling system in the preceding subsection, we should analyze the influence of sampling rates. Firstly, we consider the case of spectral sparsity level $s=1$, which means that only one frequency bin $k_{1}\in \Omega$ is occupied by the PU.

\begin{lemma} \label{lemma1}
If the numbers of samples in multiple CRs, i.e., $M_{1}, M_{2},...,M_{v}$, are different primes, and meet the requirement of
\begin{equation}
 M_iM_{j}>N, \hspace{1em} \forall \hspace{0.2em} i\neq j \in [1,v] \label{rate}
\end{equation}
then two or more CRs cannot have mirrored frequencies in the same frequency bin.
\end{lemma}

The proof of Lemma 1 is given in Appendix~\ref{appendix3}.

Secondly, considering the spectral sparsity level $s\ge2$, we find that, if the conditions in Lemma 1 are satisfied, the parameter $p$ in (\ref{bound}) is bounded by $s$. It is because only one CR can map the original frequency bin $k_j\in \Omega_i$ to the aliased frequency in $\Omega_{\textrm{A}}$, and the cardinality of the spectral support $\Omega_i$ is $s$. Therefore, we obtain the detection performance of MS$^3$ as Theorem~\ref{theorem1}.

\setcounter{theorem}{0}
\begin{theorem}
\label{theorem1}
In MS$^3$, if the numbers of samples in multiple CRs, i.e., $M_{1}, M_{2}, \cdots, M_{v}$, are different consecutive primes, and meet the requirement of $M_iM_{j}>N,~\forall~ i\neq j \in [1,v]$, using the decision rule of (\ref{decision}) the probabilities of false alarm and detection have the following bounds:
 \begin{eqnarray}
 \frac{\Gamma(Jv,\frac{\lambda_k}{2})}{\Gamma(Jv)} \le \!\!& \!\!P_{\textrm{f},k} \!\!&\!\! \le Q_{Jv} \left( \sqrt{ \frac{2}{N} \mathop{\sum}\limits_{i\in \Upsilon}^{|\Upsilon|=s} M_i \gamma_i[k] },\sqrt{\lambda_k}\right)
 \label{f1}\\
 \!\!&\!\!P_{\textrm{d}, k} \!\!& \!\!\ge Q_{Jv} \left( \sqrt{\frac{2}{N}\mathop{\sum}\limits_{i=1}^{v}M_i\gamma_i[k]},\sqrt{\lambda_k}\right). \hspace{-4em}
\label{ff2}
\end{eqnarray}
\end{theorem}

\emph{Proof}: Using (\ref{ff1}) and the bound $|\Upsilon|=p \le s$, (\ref{f1}) follows. Furthermore, when the energy of one spectral component in $\Omega$ maps to another spectral component in $\Omega$, the probability of detection will increase. Thus, the inequality of (\ref{ff2}) holds. $\Box$

\emph{Remark 1}: It can be seen from Theorem~\ref{theorem1} that the sampling rates in MS$^3$ can be much lower than the Nyquist rate because of $M_i \sim \mathcal{O}(\sqrt{N})$.
By (\ref{f1}) we note that the probability of false alarm increases when the spectral sparsity $s$ increases. In addition, the higher average sampling rate will lead to better detection performance. This is because the probability of signal overlap in the aliased spectrum can be reduced with a larger $M_i$ in each sampling channel as our discussions in Section~\ref{section2.2}. By (\ref{f1}) and (\ref{ff2}), we can see that using more sampling channels (i.e., $v$), the detection performance can be improved. It should be emphasized that there is no closed-form expression for the probabilities in Theorem~\ref{theorem1}. This is because the number of CRs that have aliased frequencies on the frequency bin $k$ cannot be predicted. Moreover, we note that the upper and lower bounds in Theorem~\ref{theorem1} can be easily computed because the Marcum-Q function can be efficiently computed using power series expansions \cite{marcum1}. Under the Neyman-Pearson criterion, we should design a test with the constraint of $P_{\textrm{f}, k}\le \alpha$. In such a scenario, we must let the upper bound of (\ref{f1}) to be $\alpha$ and solve the detection threshold $\lambda_k$ from the inverse of the Marcum-Q function. It has been shown in \cite{inverse} that the detection threshold can be calculated with low computational complexity. In addition, to calculate the detection threshold, the noise power is required to be known at the FC.

\section{Combination of Faded Signals}
\label{section3}

As shown in Section~\ref{section2}, the combining procedure in the FC is to sum up all non-faded signals at CRs and make final decisions. In this section, we investigate the combination of faded signals at CRs using the same approach as shown in (\ref{energy2}). We assume that the received primary signals at different CRs are independent and identically distributed (i.i.d.), and are faded subject to either Rayleigh or log-normal distribution.

Solving the distribution of the sum of weighted independent random variables in (\ref{energy2}) is not trivial. Hence, we use the sum of uniformly weighted random variables to approximate the sum of different weighted random variables in Theorem~\ref{theorem1}:
\begin{equation}
 \frac{2 \mathop{\sum}\limits_{i=1}^{v}{ M_i\gamma_i}}{N} \simeq \frac{2 \overline{M} }{N} \mathop{\sum}\limits_{i=1}^{v} \gamma_i= \psi \gamma_{\textrm{v}}, \hspace{2em}
 \frac{2 \mathop{\sum}\limits_{i\in \Upsilon}^{|\Upsilon|=s}{ M_i\gamma_i}}{N} \simeq \frac{2 \overline{M} }{N} \mathop{\sum}\limits_{i\in \Upsilon}^{|\Upsilon|=s} \gamma_i= \psi \gamma_{\textrm{s}} \label{rm}
\end{equation}
where $\overline{M}$ is the average $M_i$ over multiple CRs, $\psi \stackrel{\triangle}{=} \frac{2 \overline{M} }{N}$, $\gamma_{\textrm{v}} \stackrel{\triangle}{=}\sum_{i=1}^{v} \gamma_i$, and $\gamma_{\textrm{s}}\stackrel{\triangle}{=}\sum_{i\in \Upsilon}^{|\Upsilon|=s} \gamma_i$.
We note that the above approximation accuracy mainly depends on $\frac{|\overline{M}-M_i|}{N}$, where smaller $\frac{|\overline{M}-M_i|}{N}$ corresponds to more accurate approximation. Since $M_{1}, M_{2},...,M_{v}$ are chosen to be $v$ different consecutive prime numbers and the distance between primes could be very small compared to $N$, the parameter $\frac{|\overline{M}-M_i|}{N}$ will approach to zero as $N$ increases. Thus, the above approximation has little impact on the final result.

\subsection{Rayleigh Distribution}
\label{section3.1}

If the magnitudes of received signals at different CRs follow Rayleigh distribution, then the SNRs will follow exponential distribution. Hence, $\gamma_{\textrm{v}}$ and $\gamma_{\textrm{s}}$ follow Gamma distributions:
\begin{equation}
f(\gamma_{\textrm{v}})=\frac{\gamma_{\textrm{v}}^{v-1}}{\overline{\gamma}^{v}\Gamma(v)} e^{-\frac{\gamma_{\textrm{v}}}{\overline{\gamma}}}, \hspace{1em} \gamma_{\textrm{v}} \ge 0, \hspace{2em}
f(\gamma_{\textrm{s}})=\frac{\gamma_{\textrm{s}}^{s-1}}{\overline{\gamma}^{s}\Gamma(s)} e^{-\frac{\gamma_{\textrm{s}}}{\overline{\gamma}}}, \hspace{1em} \gamma_{\textrm{s}} \ge 0
\label{fr0}
\end{equation}
where $\overline{\gamma}=\mathbb{E}(\frac{|H S|^2}{\delta^2})$ denotes average SNR over multiple CRs, and $f(\cdot)$ denotes a generic probability density function (PDF) of its argument.

The average probabilities of false alarm and detection for MS$^3$ are often solved by averaging $P_{\textrm{f},k}$ in (\ref{f1}) and $P_{\textrm{d},k}$ in (\ref{ff2}) over all possible SNRs, respectively.

\begin{theorem}
\label{theorem2}
If the magnitudes of received signals at different CRs follow Rayleigh distribution, the average probabilities of false alarm ($\overline{P_{\textrm{f},k}}$) and detection ($\overline{P_{\textrm{d},k}}$) in MS$^3$ will have the following bounds
\begin{eqnarray}
\frac{\Gamma(Jv,\frac{\lambda_k}{2})}{\Gamma(Jv)} \le \!\!&\!\! \overline{P_{\textrm{f},k}} \!\!&\!\! \le \Theta ( s, Jv, \psi, \overline{\gamma}[k], \lambda_k)
\label{f3}\\
\!\!&\!\!\overline{P_{\textrm{d},k}}\!\!&\!\! \ge \Theta ( v, Jv, \psi, \overline{\gamma}[k], \lambda_k)
\label{f4}
\end{eqnarray}
where $\Theta ( x, Jv, \psi, \overline{\gamma}, \lambda)$ is defined as
\begin{equation}
\Theta=\left( 1 + \frac{\psi \overline{\gamma}}{2}\right)^{ -x} \mathop{\sum}\limits_{n=0}^{\infty} \textrm{C}_{n+x-1}^{n} \left( \frac{\psi \overline{\gamma}}{\psi \overline{\gamma}+2}\right)^n \frac{ \Gamma \left( n + Jv,\frac {\lambda}{2} \right) }{ \Gamma \left( n + Jv \right)}
\label{comp}
\end{equation}
in which $\textrm{C}_{a}^{b}$ denotes the binomial coefficient, i.e., $\textrm{C}_{a}^{b}=\frac{b!}{a!(b-a)!}$.
\end{theorem}

The proof of Theorem~\ref{theorem2} is given in Appendix~\ref{appendix4}.

\emph{Remark 2}: From Theorem~\ref{theorem2}, we can see that $0 \le \Theta \le 1$, because the term $\frac{\Gamma (a, b)}{\Gamma (a)} \in [0, 1]$ and the remaining terms can be simplified to 1. In addition, it can be proved that $\Theta$ is a monotonically increasing function with respect to $\psi$, $\overline{\gamma}$, and $x$. Therefore, both probabilities will either increase or remain the same when the average sampling rate and the average SNR increase, more sampling channels will lead to a higher probability of detection, and the average probability of false alarm can be reduced with smaller~$s$.

\emph{Remark 3}: Because (\ref{comp}) contains infinite sums, its computational complexity is directly related to the number of computed terms that are required in order to obtain a specific accuracy. As the number of computed terms, i.e., $P$, varies, the truncation error can be written as
\begin{eqnarray}
T_{\Theta}(P)&=&\left( 1 + \frac{\psi \overline{\gamma}}{2}\right)^{ -x} \mathop{\sum}\limits_{n=P}^{\infty} \textrm{C}_{n+x-1}^{n} \left( \frac{\psi \overline{\gamma}}{\psi \overline{\gamma}+2}\right)^n \frac{ \Gamma \left( n + Jv,\frac {\lambda}{2} \right) }{ \Gamma \left( n + Jv \right)} \\
&\le & \left( 1 + \frac{\psi \overline{\gamma}}{2}\right)^{ -x} \mathop{\sum}\limits_{n=P}^{\infty} \textrm{C}_{n+x-1}^{n} \left( \frac{\psi \overline{\gamma}}{\psi \overline{\gamma}+2}\right)^n \label{inq}\\
&=&1-\left( 1 + \frac{\psi \overline{\gamma}}{2}\right)^{ -x} \mathop{\sum}\limits_{n=0}^{P-1} \textrm{C}_{n+x-1}^{n} \left( \frac{\psi \overline{\gamma}}{\psi \overline{\gamma}+2}\right)^n \label{bio}
\end{eqnarray}
where the inequality of (\ref{inq}) holds because $\frac{\Gamma(n,\frac {\lambda}{2})}{\Gamma(n)}\le 1$, and (\ref{bio}) is obtained by using the binomial expansion. It can be shown that (\ref{comp}) converges very quickly. For example, in order to achieve double-precision
accuracy, only $P=30\sim40$ calculated terms are required; therefore the bounds are tractable. To solve for the detection threshold $\lambda_k$, we could use the lower bound on $P_{\textrm{f},k}$ in (\ref{f4}). This is because the lower bound can approximate $P_{\textrm{f},k}$ very well as analyzed in Appendix~\ref{appendix4} and also verified by Fig.~\ref{fig2}.

\subsection{Log-normal Distribution}
\label{section2.5}

The strength of the transmitted primary signal is also affected by shadowing from buildings, hills, and other objects. A common model is that the received power fluctuates with a log-normal distribution.
In such a scenario, the PDF of the SNR at CR~$i$, i.e., $f(\gamma_i)$, is given by
\begin{equation}
f(\gamma_i)=\frac{\xi}{\sqrt{2\pi}\sigma_i \gamma_i} \exp \left( -\frac{ \left(10 \log_{10}(\gamma_i)-\overline{\gamma}_i \right)^2}{2\sigma_i^2}\right), \hspace{1em} \gamma_i>0
\label{saa}
\end{equation}
where $\xi=10/\ln (10)$, and $\sigma_i$ (dB) denotes the standard deviation of $10 \log_{10}\gamma_i$ at CR~$i$. Note that the PDF in (\ref{saa}) can be closely approximated by a Wald distribution \cite{hongjian, wald}:
\begin{equation}
f(\gamma_i)=\sqrt{\frac{\eta_i}{2\pi}}\gamma_i^{-3/2}\exp \left(-\frac{\eta_i(\gamma_i-\theta_i)^2}{2 \theta_i^2 \gamma_i} \right), \hspace{1em} \gamma_i>0
\label{pdfg}
\end{equation}
where $\theta_i=\mathbb{E}(\gamma_i)$ denotes the expectation of $\gamma_i$, and $\eta_i$ is the shape parameter for CR~$i$. Via the method of moments, the parameters $\eta_i, \theta_i$ and $\overline{\gamma}_i$, $\sigma_i$ are related as follows:
\begin{equation}
 \theta_i = \exp \left( \frac{\overline{\gamma}_i}{\xi} + \frac{\sigma_i^2}{2\xi^2}\right), \hspace{2em}
\eta_i = \frac{\theta_i}{\exp (\frac{\sigma_i^2}{\xi^2})-1}.
\end{equation}

In the proposed system, the condition $\frac{\eta_i}{\theta_i^2}=\frac{\mathbb{E}(\gamma_i)}{\textrm{Var}(\gamma_i)}=b$ (constant) can be satisfied. Thus, $\gamma_{\textrm{s}}$ and $\gamma_{\textrm{v}}$ will also follow the Wald distribution \cite{ign2}. The PDFs of $\gamma_{\textrm{s}}$ and $\gamma_{\textrm{v}}$ are given by
\begin{equation}
f(\gamma_{\textrm{s}})=\sqrt{\frac{s\eta}{2\pi}}\gamma_{\textrm{s}}^{-3/2}\exp \left(-\frac{\eta(\gamma_{\textrm{s}}-s\theta)^2}{2s \theta^2 \gamma_{\textrm{s}}} \right), \hspace{1em} \gamma_{\textrm{s}}>0
\label{sump}
\end{equation}
\begin{equation}
f(\gamma_{\textrm{v}})=\sqrt{\frac{v\eta}{2\pi}}\gamma_{\textrm{v}}^{-3/2}\exp \left(-\frac{\eta(\gamma_{\textrm{v}}-v\theta)^2}{2v \theta^2 \gamma_{\textrm{v}}} \right), \hspace{1em} \gamma_{\textrm{v}}>0
\label{sump2}
\end{equation}
where $\eta$ and $\theta$ denote the averages of $\eta_i$ and $\theta_i$, respectively.

\begin{theorem}
\label{theorem3}
If the magnitudes of received signals at different CRs follow log-normal distribution, the average probabilities of false alarm ($\widetilde{P_{\textrm{f},k}}$) and detection ($\widetilde{P_{\textrm{d},k}}$) in MS$^3$ will be bounded as
\begin{eqnarray}
 \frac{\Gamma(Jv,\frac{\lambda_k}{2})}{\Gamma(Jv)} \le \widetilde{P_{\textrm{f},k}} \le \Lambda( s, Jv, \psi, \lambda_k, \theta[k], \eta[k])
\label{f6}\\
\widetilde{P_{\textrm{d},k}} \ge \Lambda( v, Jv, \psi, \lambda_k, \theta[k], \eta[k])
\label{f5}
\end{eqnarray}
where $\Lambda( x, Jv, \psi, \lambda, \theta, \eta)$ is defined by
\begin{equation}
\Lambda = \sqrt{\frac{2x\eta}{\pi}} e^{ \frac{\eta}{\theta} } \sum_{n=0}^{\infty} \frac{ \left( \frac{\psi}{2}\right)^n \Gamma \left( n + Jv,\frac {\lambda}{2} \right) }{ n!\Gamma \left( n + Jv \right)}\left( \sqrt{\frac{x^2\eta \theta^2}{x\psi \theta^2+\eta}}\right)^{ n-\frac{1}{2}} \textrm{K}_{n-\frac{1}{2}} \left( \sqrt{\frac{\eta (x \psi \theta^2 + \eta)}{\theta^2}} \right)
\label{slow}
\end{equation}
in which K$_{n-\frac{1}{2}}(a)$ denotes the modified Bessel function of the second kind with order $n-\frac{1}{2}$.
\end{theorem}

The proof of Theorem~\ref{theorem3} is given in Appendix~\ref{appendix5}.

\emph{Remark 4}: Because (\ref{slow}) contains infinite sums, the truncation error $T_{\Lambda} (P)$ must be considered. Similar to (\ref{inq}), the truncation error can be written as
\begin{eqnarray}
 T_{\Lambda}(P) & \le & \sqrt{ \frac{2x\eta}{\pi}} e^{\frac{\eta}{\theta}} \mathop{\sum}\limits_{n=P}^{\infty} \frac{ \left( \frac{\psi}{2} \right)^n \left( \sqrt{\frac{x^2\eta\theta^2}{x\psi\theta^2 + \eta}}\right)^{ n - \frac{1}{2}}}{n!} \textrm{K}_{n - \frac{1}{2}} \left( \frac{\sqrt{\eta(x \psi \theta^2 + \eta)}}{\theta^2} \right) \hspace{1em}\nonumber\\
 & = & 1 -\sqrt{ \frac{2x\eta}{\pi}} e^{\frac{\eta}{\theta}} \mathop{\sum}\limits_{n=0}^{P-1} \frac{ \left( \frac{\psi}{2} \right)^n \left( \sqrt{\frac{x^2\eta\theta^2}{x\psi\theta^2 + \eta}}\right)^{ n - \frac{1}{2}}}{n!} \textrm{K}_{n - \frac{1}{2}} \left( \frac{\sqrt{\eta(x \psi \theta^2 + \eta)}}{\theta^2} \right). \hspace{1em}
 \label{trun}
\end{eqnarray}
It can be shown that (\ref{trun}) decreases to zero very quickly as $P$ increases, therefore the bounds in Theorem~\ref{theorem3} are easy to compute.

\section{Simulation results}
\label{section4}

In our simulations, we assume that the CRs are organized as shown in Fig.~\ref{fig1} and adopt the following configurations unless otherwise stated.  We use the wideband analog signal model in \cite{was1} and thus the received signal $x_{\textrm{c},i}(t)$ at CR $i$ has the form:
\begin{equation}
x_{\textrm{c},i}(t) = \mathop{\sum}\limits_{l=1}^{N_b} |H_{i,l}|\sqrt{E_{l}} B_l \cdot \textrm{sinc} \left(B_l(t - \Delta)\right) \cdot \cos \left(2\pi f_{l} (t - \Delta) \right)+z(t)
\end{equation}
where sinc$(x)=\frac{\sin (\pi x)}{\pi x}$, $\Delta$ denotes a random time offset, $z(t)$ is AWGN, i.e., $z(t)\sim \mathcal{N}(0,1)$, $E_{l}$ is the transmit power at PU, and $H_{i,l}$ denotes the discrete frequency response between the PU and CR $i$ in subband $l$. We generate $v=22$ independent channels according to the fading environment, and regenerate them for next observation time. The received signal $x_{\textrm{c},i}(t)$ consists of $N_b=6$ non-overlapping subbands. The $l$-th subband is in the frequency range of [$f_{l}-\frac{B_l}{2}$, $f_{l}+\frac{B_l}{2}$], where the bandwidth $B_l=1\sim 10$~MHz and $f_{l}$ denotes the center frequency. The center frequency of the subband $l$ is randomly located within $[\frac{B_l}{2}, W-\frac{B_l}{2}]$ (i.e., $f_{l} \in [\frac{B_l}{2}, W-\frac{B_l}{2}]$), where the overall signal bandwidth $W=10$ GHz. If the wideband signal were sampled at the Nyquist rate $f=2W$ for $T=20$~$\mu$s, after segment division with $J=5$, the number of Nyquist samples per segment would be $N=80,000$; thus, using FFT-based approach, the frequency resolution is $\frac{1}{T/J}=0.25$ MHz.  In MS$^3$, the received signal is sampled by using different sub-Nyquist rates at different CRs. To be specific, the numbers of samples in multiple CRs are chosen by using Theorem 1 and we choose the first prime $M_1 \approx a\sqrt{N}$ ($a \ge 1$) and its $v-1$ neighboring and consecutive primes. The spectral observations are obtained by applying an FFT to these sub-Nyquist samples in each channel. \textbf{Then the signal energy is calculated in the spectral domain using (\ref{sum}), and the energy vectors are transmitted from the CRs to the FC using dedicated common control channels. These channels are assumed to be Rayleigh block fading channels (constant channel gains over one time block) corrupted by circularly symmetric complex
Gaussian noise with zero mean and unit variance. In addition, we consider that the channel power gain of the common control channel is normalized to unit and the average SNR as received at the FC is 15 dB.} In the FC, we form the test statistic by using (\ref{energy2}). We define the compression rate as the ratio between the number of samples at the sub-Nyquist rate and the number of samples at the Nyquist rate, i.e., $\frac{M}{N}$ where $M$ denotes the average number of sub-Nyquist samples at CRs.  Spectrum sensing results are obtained by using the decision rule (\ref{decision}) and varying the detection threshold $\lambda_k$.

\begin{figure}[!t]
\centering
\centerline{\includegraphics[width=5in]{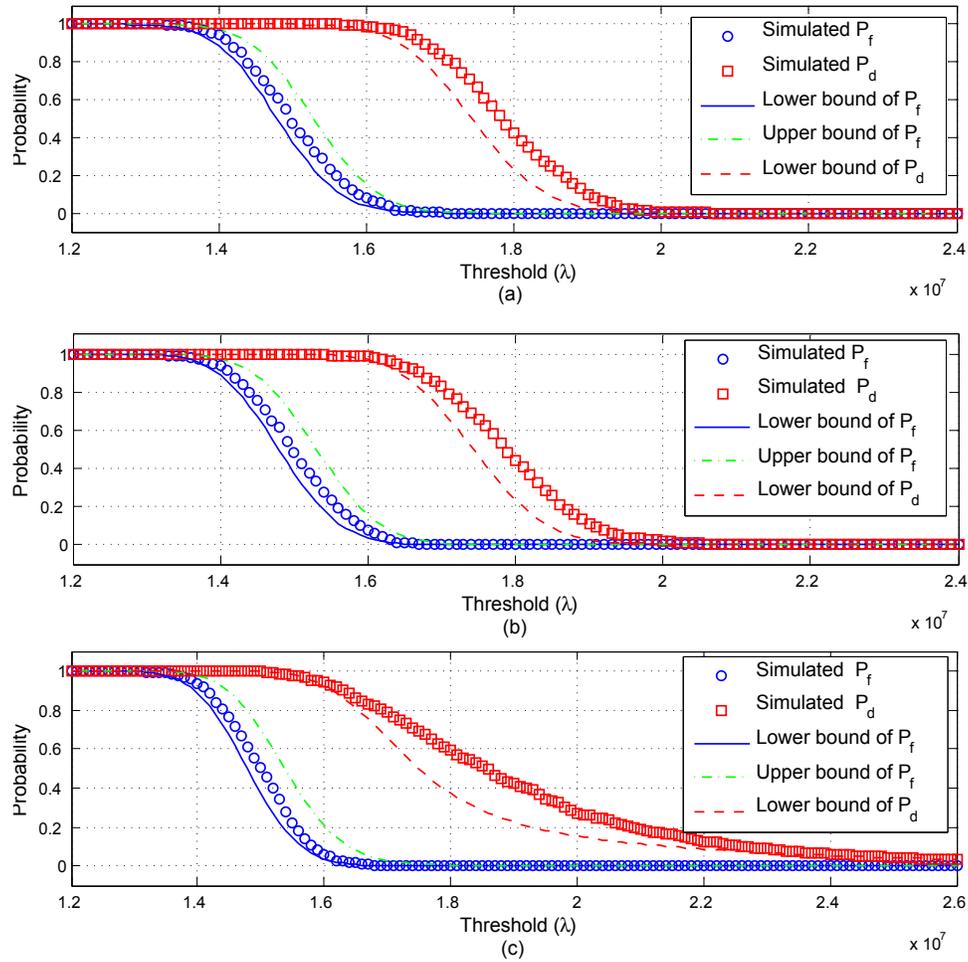}}
\vspace{-0.5em}
\caption{Comparisons of simulation results and analytical results for the probabilities of false alarm and detection when MS$^3$ combining (a)non-faded signals, (b) Rayleigh faded signals, and (c) Log-normal shadowed signals with the received SNR$=5$ dB (at CRs) and $\sigma=4$ dB.}
\vspace{-0.5em}
\label{fig2}
\end{figure}

In Fig.~\ref{fig2}, we verify the theoretical results in (\ref{f1})-(\ref{ff2}), (\ref{f3})-(\ref{f4}), and (\ref{f6})-(\ref{f5}) by comparing them with the simulated results. It shows that the lower bound on the probability of false alarm can tightly predict the simulated results while the upper bound seems relatively loose. It is because that the assumption (i.e., all $s$ components in the Nyquist DFT spectrum will be mapped to the same location when the signal is sub-Nyquist sampled) for deriving the upper bound can rarely occur. Fig.~\ref{fig2} also illustrates that the lower bound on the probability of detection can successfully predict the trend of simulated results. Comparing the faded signal cases with the non-faded signal case, it is found that, when combining faded signals, the probability of detection declines more slow than that combining non-faded signals. This is more obvious for the case of combining signals following log-normal distribution as shown in Fig.~\ref{fig2}(c).

\begin{figure}[!pt]
\centerline{\includegraphics[height=3.6in]{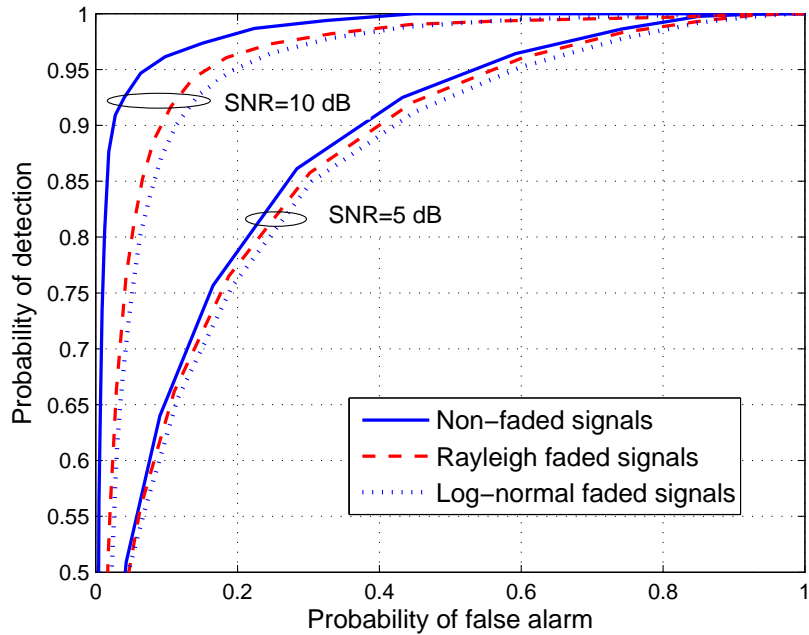}}
\caption{Receiver operating characteristic curves of MS$^3$ for combining non-faded signals or faded signals when the compression rate $\frac{M}{N}=0.0219$ and the number of segments $J=5$. The wideband signal is observed by $22$ CRs at different sampling rates (the average sampling rate is $448.68$ MHz). }
\label{fig5}
\end{figure}

\begin{figure}[!t]
\centering
\centerline{\includegraphics[height=3.6in]{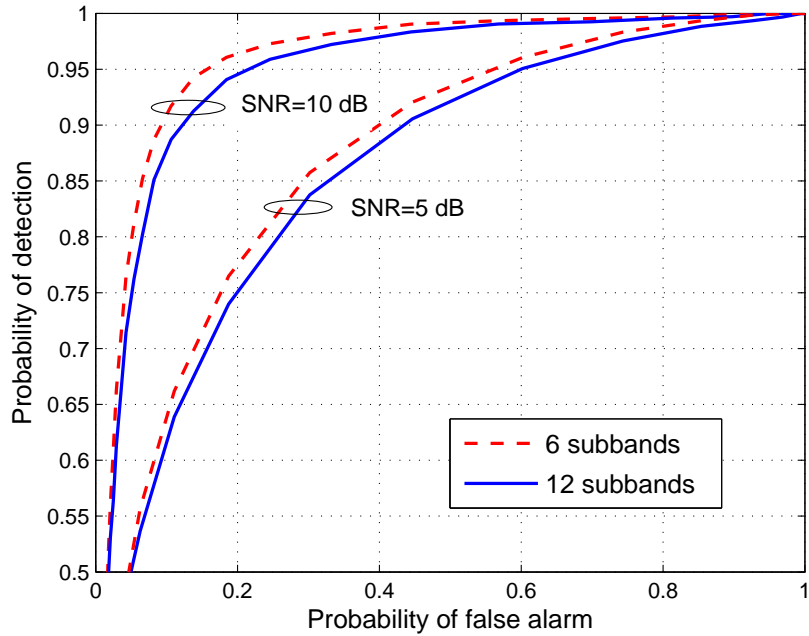}}
\vspace{-0.9em}
\caption{The performance of MS$^3$ for combining Rayleigh faded signals with $v=22$ and $\frac{M}{N}=0.0219$, when the received SNR at CRs and the number of subbands change.}
\label{fig4}
\end{figure}

Fig.~\ref{fig5} shows the receiver operating characteristic (ROC) curves of MS$^3$ when combining non-faded and faded signals. When the average SNR as received at CRs is 5~dB, the performance of MS$^3$ combining faded signals is roughly the same as that of combining non-faded signals. This is because the strength of the signal is mostly masked by the noise. In contrast, the detection performance of MS$^3$ combining non-faded signals outperforms that of combining faded signals when SNR=10 dB. In addition, it is seen that the performance of MS$^3$ combining log-normal shadowed signals is the poorest. Nonetheless, even for log-normal shadowed signals, MS$^3$ has a probability of nearly $90\%$ for detecting the presence of PUs when the probability of false alarm is $10\%$, with the compression rate of $\frac{M}{N}=0.0219$. To investigate the influence of $s$ and SNR, we use Fig.~\ref{fig4} to show the performance of MS$^3$ when the received signals are faded according to Rayleigh distribution with different values of $s$ (proportional to the number of subbands). We see that, as the number of subbands decreases, the detection performance improves for the same SNR. The performance improvement of MS$^3$ stems from that, for a fixed number of sampling channels, decreasing $s$ makes it easier to distinguish the occupied frequencies from the aliased frequencies as discussions in Section \ref{section2.3}.

\begin{figure}[!h]
\begin{center}$
\begin{array}{cc}
\includegraphics[width=3.5in]{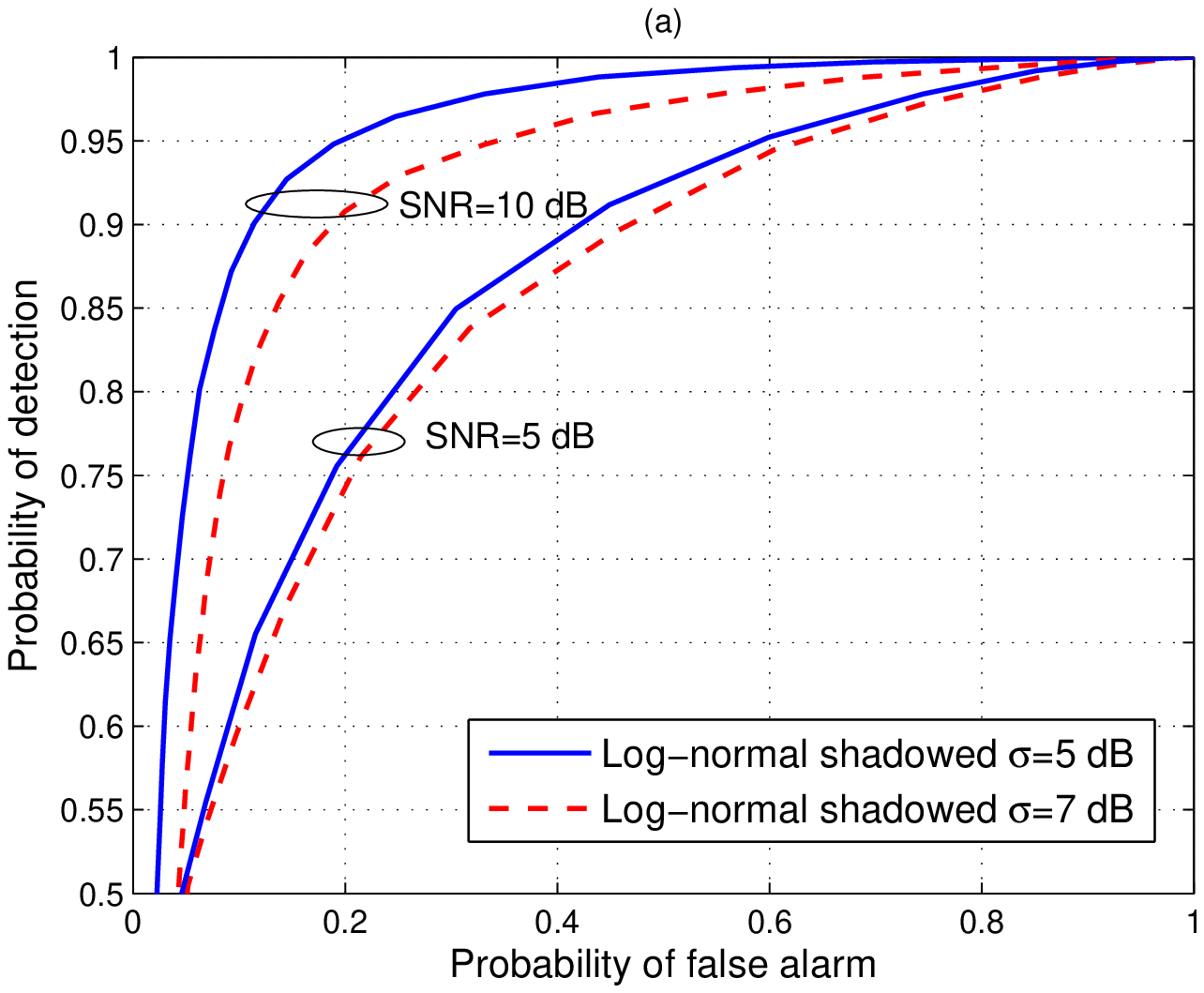} &
\includegraphics[width=3.5in]{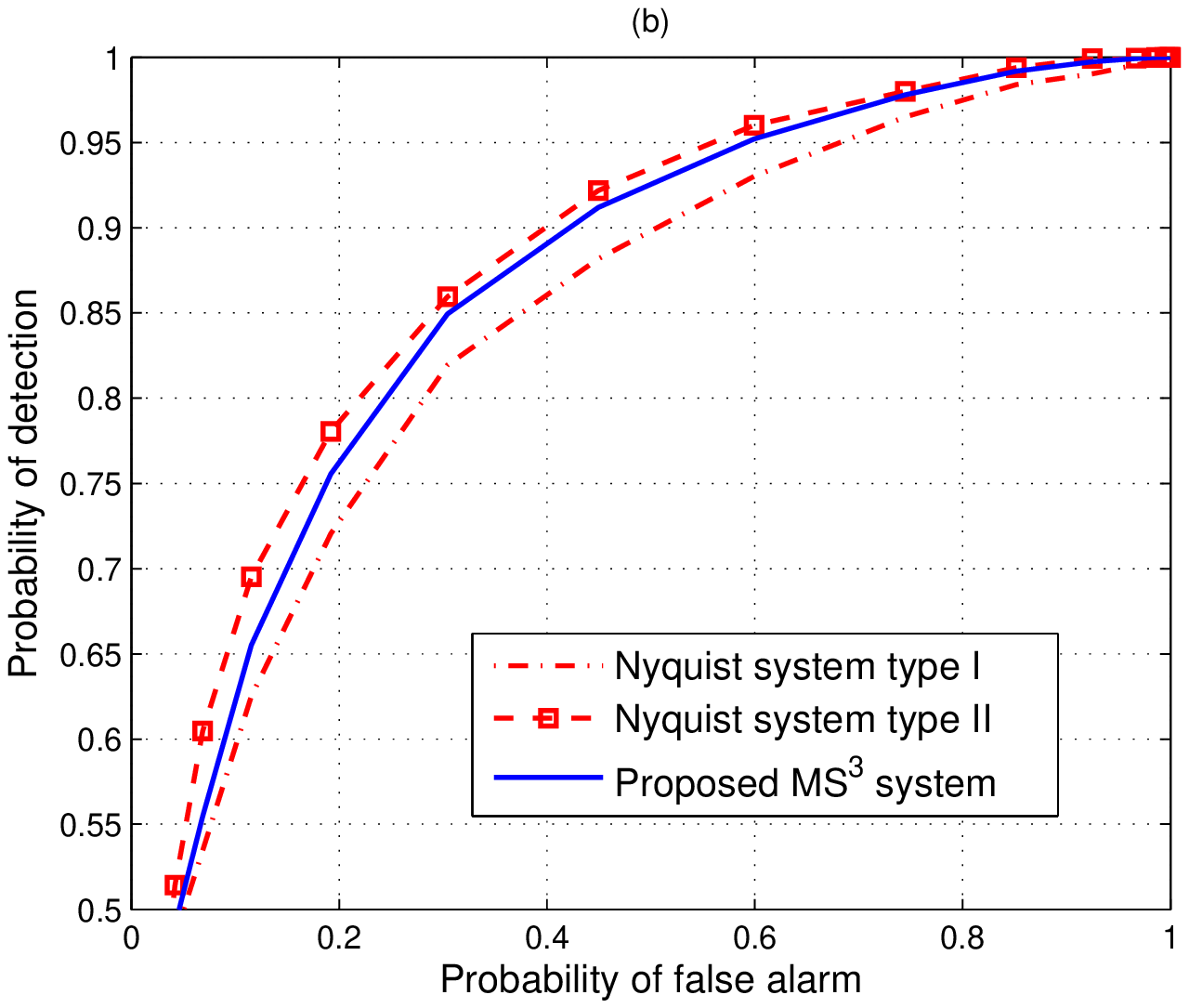}
\end{array}$
\end{center}
\vspace{-0.8em}
\caption{Receiver operating characteristic curves for combining log-normal faded signals: (a) MS$^3$ when the standard deviation, i.e., $\sigma$, and the average SNR at CRs vary, (b) comparison between MS$^3$ and Nyquist systems when $\sigma=5$ dB and the average SNR as received at CRs is $5$ dB. In simulations, the wideband signal is sampled at different sampling rates by $22$ ADCs with the average sampling rate of $448.68$ MHz, and the compression rate is $\frac{M}{N}=0.0219$ and $J=5$.}
\label{fig6}
\end{figure}

Fig.~\ref{fig6}(a) depicts the influence of the standard deviation when the MS$^3$ system combines log-normal shadowed signals. It can be seen that a larger standard deviation will lead to worse detection performance for the MS$^3$ system. It is because a larger $\sigma$ is equivalent to a longer tail in the log-normal distribution, thus making the detection more difficult. \textbf{In Fig.~\ref{fig6}(b), we compare the performance of MS$^3$ with that of Nyquist systems. In the Nyquist system type~I, each CR is given an orthogonal subband (wideband spectrum is divided into several equal-length subbands) to sense using Nyquist rate, while their decisions are sent back to the FC. In the Nyquist system type~II, we assume that each CR must sense all wideband spectrum non-cooperatively, thus requiring multiple standard ADCs in each node to cover all wideband spectrum. After signal sampling, all measurements are sent back to the FC, where equal gain combining approach is adopted to fuse data and then energy detection is used for spectrum sensing. Fig.~\ref{fig6}(b) shows that the proposed system has superior performance to the Nyquist system type~I, but inferior performance to the Nyquist system type~II. The poor performance of the Nyquist system type~I mainly results from the lack of spatial diversity gain. In the Nyquist system type~I, each subband is only sensed by one CR as each CR is given an orthogonal subband to sense, which cannot take advantage of spatial diversity. In contrast, both the proposed system and the Nyquist system type~II are monitoring each subband using several CRs, thus taking advantage of spatial diversity. It can also be seen that the Nyquist system type~II has marginal performance gain over the proposed system, however, at the expense of much higher implementation complexity as discussed below.}

In Table~\ref{table:compare2}, we compare the implementation complexity of MS$^3$ with that of the Nyquist systems, when the received signals at different CRs are faded according to Rayleigh distribution.  Here, we consider the comparison metric: the number of same-sampling-rate ADCs for achieving $P_{\textrm{d}}\ge90\%$ and $P_{\textrm{f}}\le10\%$, because practical CRs often have requirements on the probabilities of detection and false alarm to secure the performance of both CRs and PUs.
We can see that, when there exist 10 CRs, MS$^3$ requires each CR equipped with a single ADC with an average sampling rate of $957.54$~MHz; thus, the whole CR network only requires 10 low-rate ADCs.
In contrast, the Nyquist system type~I requires 21 ADCs in total, because of $21\times 957.54$ MHz$\approx 20$ GHz for covering $10$ GHz spectrum based on Nyquist sampling theorem.
In the Nyquist system type~II, $210$ ADCs (with the average sampling rate $957.54$~MHz) will be required because each CR will require 21 ADCs similar to the Nyquist system type~I. \textbf{Thus, the system complexity of MS$^3$ is approximately half of that of the Nyquist system type~I and much less than that of the Nyquist system type~II. }

\begin{figure}[!h]
\begin{center}$
\begin{array}{cc}
\includegraphics[width=3.5in]{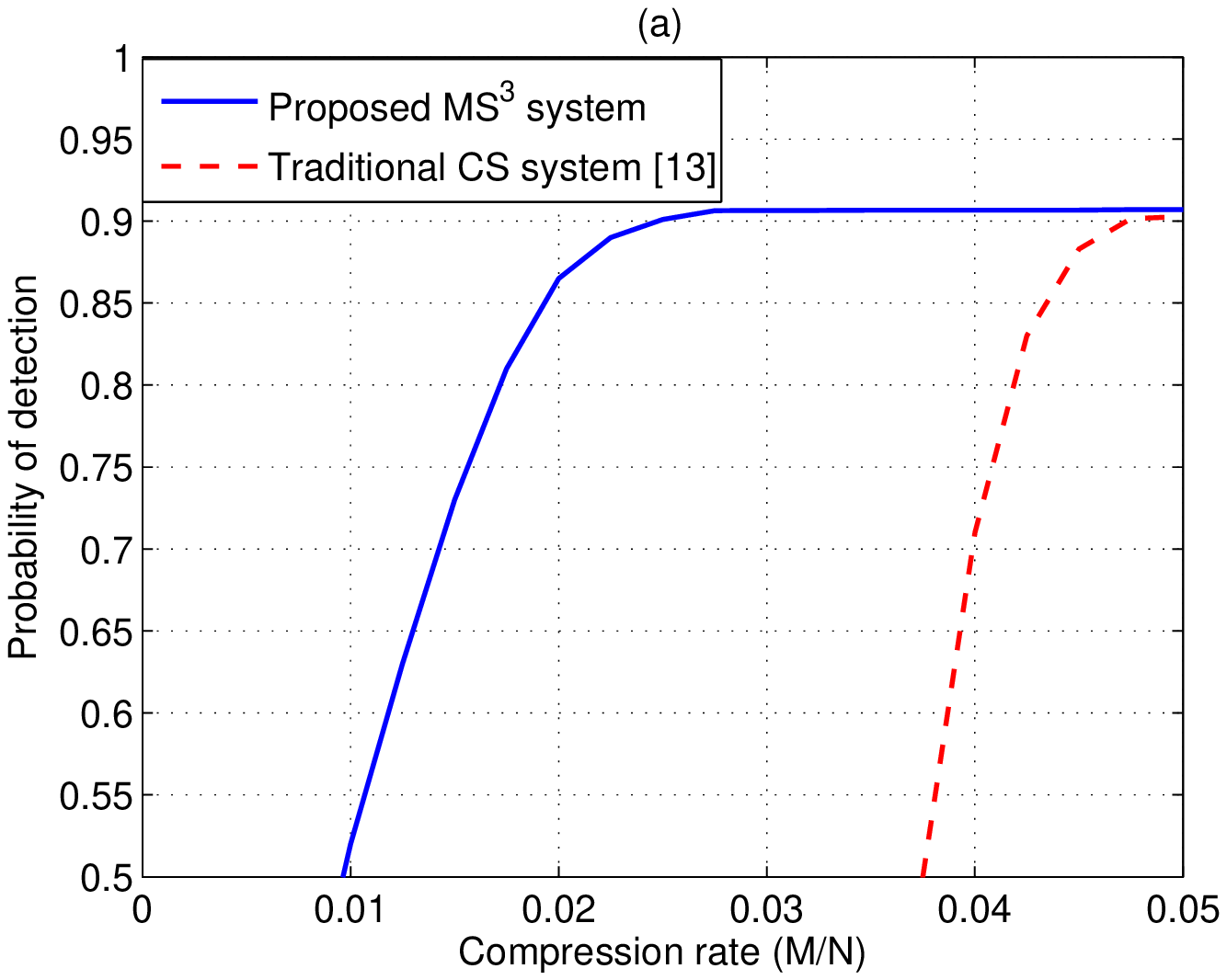} &
\includegraphics[width=3.5in]{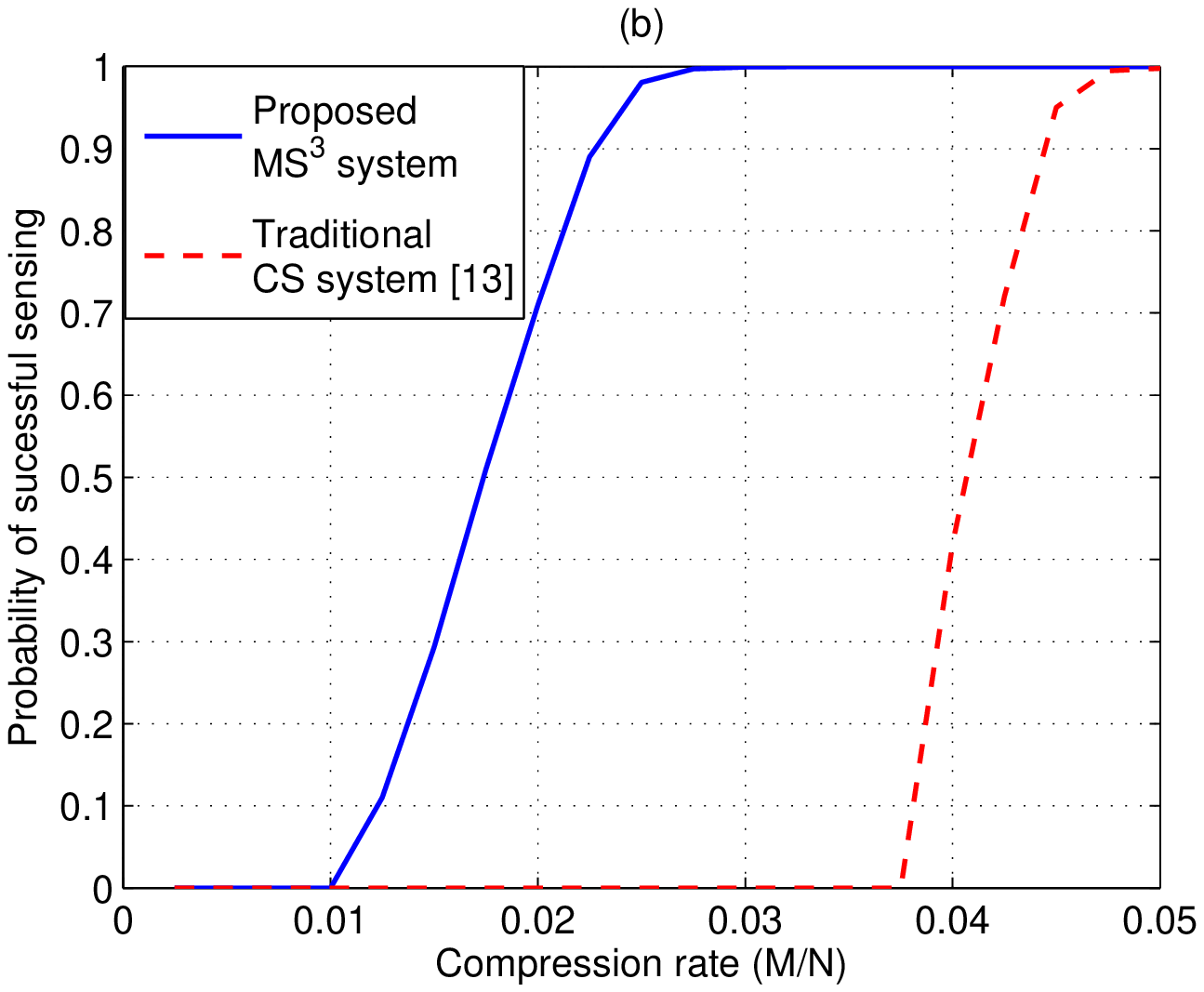}
\end{array}$
\end{center}
\vspace{-0.8em}
\caption{Comparison between MS$^3$ and CS-based system \cite{scs1}: (a) the probability of detection when the probability of false alarm is set to $10\%$, and (b) the probability of successful sensing which is defined as the probability of achieving both $P_{\textrm{d}}\ge90\%$ and $P_{\textrm{f}}\le 10\%$. In simulations, the average SNR as received at CRs is $10$ dB and the number of CRs is $v=22$. }
\label{fig7}
\end{figure}

In Fig.~\ref{fig7}, we choose the CS-based system in \cite{scs1} as a benchmark system due to its high impact and outstanding performance. The comparison between the proposed MS$^3$ system and the benchmark system is provided. We assume that $v=22$ CRs are collaborating for wideband spectrum sensing in both systems, in order to increase the reliability of spectrum sensing by exploiting spatial diversity. We can see from Fig.~\ref{fig7}(a) that MS$^3$ outperforms the CS-based system for every compression rate. \textbf{ In Fig.~\ref{fig7}(b), it is seen that, compared with the benchmark system, MS$^3$ has better compression capability. Using MS$^3$, the probability of successful sensing becomes larger than $90\%$ when the compression rate $\frac{M}{N} \ge 0.023$. In contrast, the benchmark system can achieve the probability of successful sensing $90\%$ only when the compression rate $\frac{M}{N} \ge 0.045$.} Furthermore, as shown in Table~\ref{table1}, we can find that the computational complexity of MS$^3$ is $\mathcal{O}(N \log N)$ due to the energy detection with FFT operations, rather than $\mathcal{O}\left(N (M+\log N)\right)$ in the CS-based system, where $M$ is usually much larger than $\log N$. The complexity of the CS-based system is caused by both the matrix multiplication operations and the FFT operations for spectral recovery. To sum up, with the same computational resources, MS$^3$ has a relatively smaller spectrum sensing overhead than the CS-based system, not only because of the better compression capability (less data transmission results in shorter transmission time), but also due to the lower computational complexity.

\section{Conclusions}
\label{section5}

In this paper, we have presented a novel system, i.e., MS$^3$, for wideband spectrum sensing in CR networks. MS$^3$ can relax the wideband spectrum sensing requirements of CRs due to its capability of sub-Nyquist sampling. It has been shown that, using sub-Nyquist samples, the wideband spectrum can be sensed in a collaborative manner without spectral recovery, leading to a high energy-efficiency and a low spectrum sensing overhead. Moreover, we have derived closed-form bounds for the performance of MS$^3$ when combining faded or shadowed signals.

Simulation results have verified the derived bounds on the probabilities of false alarm and detection. It has also been shown that using partial measurements, MS$^3$ has superior performance even under low SNR scenarios. The performance of MS$^3$ improves as either the number of CRs or the average sampling rate increases. Compared to the existing wideband spectrum sensing methods, MS$^3$ not only provides computation and memory savings, but also reduces the hardware acquisition requirements and the energy costs at CRs. 

\appendices

\section{Relationship between Nyquist DFT Spectrum and Sub-Nyquist DFT Spectrum}
\label{appendix1}
Using the Poisson summation formula \cite{poi}, (\ref{x1}), and (\ref{sspectrum}), we obtain:
\begin{equation}
f_i\sum_{l \in \mathbb{Z}} X_{\textrm{c},i}(w+ f_i l)=\sum_{n \in \mathbb{Z}}y_i[n] e^{-\jmath 2 \pi w n}=\sum_{n =0}^{M_i-1}y_i[n] e^{-\jmath 2 \pi w n}=Y_i(w)
\label{link1}
\end{equation}
where $X_{\textrm{c},i}(w)=\int_{-\infty}^{\infty} x_{\textrm{c},i}(t) e^{-\jmath 2\pi w t} dt$. Similar to (\ref{link1}), by using (\ref{x0}) and (\ref{spectrum}), we can obtain:
\begin{equation}
f\sum_{l \in \mathbb{Z}} X_{\textrm{c},i}(w+ f l)=\sum_{n \in \mathbb{Z}}x_i[n] e^{-\jmath 2 \pi w n}=\sum_{n =0}^{N-1}x_i[n] e^{-\jmath 2 \pi w n}=X_i(w).
\label{link2}
\end{equation}
As the received signal is bandlimited and $f \ge 2W$, $X_i(w)=f X_{\textrm{c},i}(w)$ holds for $w \in [-\frac{W}{2}, \frac{W}{2}]$. Substituting it to (\ref{link1}), we obtain $Y_i(w)=\frac{f_i}{f} \sum_{l=-\infty}^{\infty} X_i(w+ f_i l)$. In a discrete form, we end up with:
\begin{equation}
 Y_{i}[m]=\frac{M_i}{N} \mathop{\sum}\limits_{l=-\infty}^{\infty} X_{i} [m+lM_i], \;\;\; m=0, 1, \cdots, M_i-1.
\label{link3}
\end{equation}

\section{Probability of Signal Overlap at Sub-Nyquist Sampling}
\label{appendix2}

As $s$ spectral components are distributed over the frequency bins of $0,1,\cdots, N-1$, the probability of the frequency bin $k$ belonging to the spectral support $\Omega$ is $P=\Pr (k \in \Omega)=\frac{s}{N}$. Let $q$ denote the number of spectral components overlapped on the frequency bin $m$, using (\ref{su3}) the probability of no signal overlap is given by
\begin{equation}
\Pr(q < 2 ) = \Pr(q=0)\!+\!\Pr(q=1) = (1\!-\!P)^{\lceil \frac{N}{M_i} \rceil}\!+ \!\binom{\lceil \frac{N}{M_i} \rceil}{1} P (1\!-\!P)^{\lceil \frac{N}{M_i} \rceil\!-\!1}
\label{jinsi}
\end{equation}
where $\lceil \frac{N}{M_i} \rceil$ denotes the number of summations in (\ref{su3}).
Substituting $P=\frac{s}{N}$ into (\ref{jinsi}) while choosing sub-Nyquist sampling rate in MS$^3$ such that $M_i=\sqrt{N}$, we obtain
\begin{equation}
\Pr(q < 2 ) = \left(\frac{N\!-\!s}{N}\right)^{\frac{N}{M_i}}\!+\! \frac{s}{M_i}\left(\frac{N\!-\!s}{N}\right)^{\frac{N\!-\!M_i}{M_i}} \!= \!\frac{(\frac{N\!-s}{N})^{\sqrt{N}}(N\!-\!s\!+\!s\sqrt{N})}{N\!-\!s}.
\label{jinsi1}
\end{equation}
It can be tested that $\Pr(q < 2 )$ approaches to 1 when choosing $N$ such that $N \gg s$. Thus the probability of signal overlap approaches to zero under the condition we choose.

\section{Proof of Lemma 1}
\label{appendix3}

Let $M_i$ and $M_{j}$ denote the number of samples at CRs $i$ and $j$, respectively. Using (\ref{fold}) and (\ref{mirror}), we can represent the aliased frequencies projected from $k_{1}\in \Omega$ by
\begin{eqnarray}
 g_{i} & = & |k_{1}|_{ \bmod( M_i )} + lM_i=k_{1} - hM_i + lM_i, \; h\neq l \\
g_{j} & = & |k_{1}|_{ \bmod( M_{j} )} + \check{l}M_{j} = k_{1} - \check{h}M_{j} + \check{l}M_{j}, \; \check{h}\neq \check{l}
\end{eqnarray}
where integers $h$ and $\check{h}$ are quotients from modulo operations, and $l-h \in [-\lceil \frac{N}{M_i} \rceil+1, \lceil \frac{N}{M_i} \rceil-1]$, $\check{l}-\check{h} \in [-\lceil \frac{N}{M_{j}} \rceil+1, \lceil \frac{N}{M_{j}} \rceil-1]$, in which $\lceil \frac{N}{M_i} \rceil$ gives the smallest integer not less than $\frac{N}{M_i}$.

Avoiding $g_{i}=g_{j}$ is equivalent to avoiding $(l-h)M_i=(\check{l}-\check{h})M_{j}$. If $M_i$ and $M_{j}$ are different primes, the condition $\max(|l-h|)<M_{j}$ (i.e., $\lceil \frac{N}{M_i} \rceil-1<M_{j}$) will satisfy this. After simplification, the condition $M_iM_{j}>N$ is obtained. Moreover, if it holds for any two CRs, the case for more than two CRs will also hold.

\section{Proof of Theorem~~\ref{theorem2}}
\label{appendix4}

If the received signals at CRs are Rayleigh faded, the lower bound on the average probability of false alarm will remain as it is independent of the SNR. Using (\ref{f1}), (\ref{rm}), and (\ref{fr0}), the upper bound on the average probability of false alarm can be calculated by
\begin{equation}
\overline{P_{\textrm{f},k}}^{\textrm{up}}= \int_{0}^{\infty} Q_{Jv} \hspace{-0.2em} \left( \sqrt{\psi \gamma_{\textrm{s}}},\sqrt{\lambda_k}\right) \frac{\gamma_{\textrm{s}}^{s-1}}{\overline{\gamma}^{s}\Gamma(s)} e^{-\frac{\gamma_{\textrm{s}}}{\overline{\gamma}}} d \gamma_{\textrm{s}}.
\label{upperb}
\end{equation}
Rewriting the Marcum Q-function by using (4.74) in \cite{overfading} and (8.352-2) in \cite{tablei}, we obtain:
\begin{equation}
Q_{Jv} \left( \sqrt{\psi \gamma_{\textrm{s}}},\sqrt{\lambda_k}\right) = \mathop{\sum}\limits_{n=0}^{\infty} \frac{ \left( \frac{\psi \gamma_{\textrm{s}}}{2} \right)^n e^{-\frac{\psi \gamma_{\textrm{s}}}{2}} }{n!} \frac{\Gamma (n + Jv, \frac{\lambda_k}{2} )}{\Gamma (n + Jv)}.
\label{anoth}
\end{equation}
Substituting (\ref{anoth}) into (\ref{upperb}), we can rewrite (\ref{upperb}) as
\begin{equation}
\overline{P_{\textrm{f},k}}^{\textrm{up}} = \frac{1}{\overline{\gamma}^s } \mathop{\sum}\limits_{n=0}^{\infty} \frac{ \left( \frac{\psi }{2}\right)^n \Gamma (n + Jv, \frac{\lambda_k}{2} )}{n! (s - 1)!\Gamma (n + Jv)} \int_{0}^{\infty} \gamma_{\textrm{s}}^{n+s-1} e^{ -\frac{\psi \gamma_{\textrm{s}}}{2}- \frac{\gamma_{\textrm{s}}}{\overline{\gamma}} } d \gamma_{\textrm{s}}.
\end{equation}
Calculating the integral by using (3.351-3) in \cite{tablei}, we end up with
\begin{equation}
\overline{P_{\textrm{f},k}}^{\textrm{up}} = \left( 1 + \frac{\psi \overline{\gamma}}{2}\right)^{-s} \mathop{\sum}\limits_{n=0}^{\infty} \textrm{C}_{n+s-1}^{n} \left( \frac{\psi \overline{\gamma}}{\psi \overline{\gamma}+2}\right)^{n} \frac{ \Gamma \left( n + Jv,\frac {\lambda_k}{2} \right) }{ \Gamma \left( n + Jv \right)}.
\end{equation}
Similarly, we can obtain the lower bound on the average probability of detection.

\section{Proof of Theorem~\ref{theorem3}}
\label{appendix5}

If the received signals are shadowed according to log-normal distribution, the lower bound on $\widetilde{P_{\textrm{f},k}}$ in (\ref{f6}) will remain. By (\ref{sump}), the upper bound on the probability of false alarm can be given by
\begin{equation}
\widetilde{P_{\textrm{f},k}}^u= \int_{0}^{\infty} Q_{Jv} \hspace{-0.2em} \left( \sqrt{\psi \gamma_{\textrm{s}}},\sqrt{\lambda_k}\right) \sqrt{\frac{s\eta}{2\pi}}\gamma_{\textrm{s}}^{-3/2}\exp \left(-\frac{\eta(\gamma_{\textrm{s}}-s\theta)^2}{2s \theta^2 \gamma_{\textrm{s}}} \right) d \gamma_{\textrm{s}}.
\label{upperbo}
\end{equation}
Substituting (\ref{anoth}) into (\ref{upperbo}), we calculate $\widetilde{P_{\textrm{f},k}}^u$ as
\begin{equation}
\widetilde{P_{\textrm{f},k}}^u = \sqrt{\frac{s\eta}{2\pi}} \sum_{n=0}^{\infty} \frac{ \left( \frac{\psi }{2}\right)^n \Gamma \left( n + Jv,\frac {\lambda_k}{2} \right) }{n!\Gamma \left( n + Jv \right)} \int_{0}^{\infty} \gamma_{\textrm{s}}^{n-\frac{3}{2}} \exp \left( -\frac{s\psi \theta^2 + \eta}{2s\theta^2}\gamma_{\textrm{s}} - \frac{s\eta}{2\gamma_{\textrm{s}}} + \frac{\eta}{\theta} \right) d \gamma_{\textrm{s}}.
\label{lala}
\end{equation}
Using (3.471-9) in \cite{tablei} for calculating the integral in (\ref{lala}), we obtain
\begin{equation}
\widetilde{P_{\textrm{f},k}}^u = \sqrt{\frac{2s\eta}{\pi}} e^{ \frac{\eta}{\theta} } \sum_{n=0}^{\infty} \frac{ \left( \frac{\psi}{2}\right)^n \Gamma \left( n + Jv,\frac {\lambda_k}{2} \right) }{ n!\Gamma \left( n + Jv \right)} \left( \sqrt{ \frac{s^2\eta \theta^2}{s\psi \theta^2 + \eta}}\right)^{ n-\frac{1}{2}} \textrm{K}_{n-\frac{1}{2}} \left( \sqrt{\frac{\eta (s \psi \theta^2 + \eta)}{\theta^2}} \right).
\end{equation}
Likewise, the lower bound on the average probability of detection can be approximated.


\ifCLASSOPTIONcaptionsoff
 \newpage
\fi

\bibliographystyle{IEEEtran}

\bibliography{cwpf}


\begin{table}[!t]
\centering
\caption{Implementation complexity comparison of MS$^3$ and the Nyquist systems when the received signals are faded according to Rayleigh distribution with ten Decibel received SNR at CRs.} 
\begin{tabular}{l || l | l | l | l} 
\hline 
Number of CRs for & \raisebox{-2ex}{10} & \raisebox{-2ex}{20} & \raisebox{-2ex}{30} & \raisebox{-2ex}{40} \\
 Wideband Spectrum Sensing ($v$) & & & & \\
\hline 
Required Number of ADCs & \raisebox{-2ex}{10} & \raisebox{-2ex}{20} & \raisebox{-2ex}{30} & \raisebox{-2ex}{40} \\
in the Proposed MS$^3$ System & & & & \\
\hline
Required Number of ADCs in & \raisebox{-2ex}{21} & \raisebox{-2ex}{40} & \raisebox{-2ex}{58} & \raisebox{-2ex}{74} \\
the Nyquist System Type I & & & & \\
\hline
Required Number of ADCs in & \raisebox{-2ex}{21$\times$10} & \raisebox{-2ex}{40$\times$20} & \raisebox{-2ex}{58$\times$30} & \raisebox{-2ex}{74$\times$40} \\
the Nyquist System Type II & & & & \\
\hline
Average Sampling Rate of ADCs & \raisebox{-2ex}{957.54} & \raisebox{-2ex}{513.08} & \raisebox{-2ex}{350.34} & \raisebox{-2ex}{276.77} \\
 in the Above Systems ($\overline{f}$ in MHz) & & & & \\
\hline
\raisebox{-1ex}{Sample Reduction Rate I}& \raisebox{-2ex}{47.62\%} & \raisebox{-2ex}{50\%} & \raisebox{-2ex}{51.72\%} & \raisebox{-2ex}{54.05\%} \\
($\frac{\textrm{Total samples in MS$^3$}}{\textrm{Total samples in Nyquist type I}}$) & & & & \\
\hline
\raisebox{-1ex}{Sample Reduction Rate II}& \raisebox{-2ex}{4.76\%} & \raisebox{-2ex}{2.5\%} & \raisebox{-2ex}{1.72\%} & \raisebox{-2ex}{1.35\%} \\
($\frac{\textrm{Total samples in MS$^3$}}{\textrm{Total samples in Nyquist type II}}$) & & & & \\
\hline
\end{tabular}
\label{table:compare2}
\end{table}

\begin{table}[!ht]
\caption{Comparisons of wideband spectrum sensing techniques.} 
\vspace{-0.6em}
\centering 
\begin{tabular}{c c c c c} 
\hline 
\raisebox{-2.5ex}{Approach} & \raisebox{-1ex}{Compression} & \raisebox{-1ex}{ADC/DSP} & \raisebox{-1ex}{Implementation} & \raisebox{-1ex}{Computational} \\ [0.5ex]
& \raisebox{1ex}{Capability} & \raisebox{1ex}{Type} & \raisebox{1ex}{Complexity} & \raisebox{1ex}{Complexity} \\
\hline 
Wavelet detection & $\times$ & Nyquist & low &
$\mathcal{O}(N\log N)$ \vspace{0.6em}\\
Multiband joint detection & $\times$ & Nyquist & high & $\mathcal{O}\left(N\log N \right)$ \vspace{0.6em}\\
CS-based detection & $\surd$ & sub-Nyquist & medium &$\mathcal{O}\left(N(M+ \log N) \right)$ \vspace{0.6em}\\
Proposed system & $\surd$ & sub-Nyquist & low & $\mathcal{O}(N\log N)$ \vspace{0.6em} \\ \hline
\end{tabular}
\vspace{-0.6em}
\label{table1} 
\end{table}

\end{document}